\documentclass[aps, prx, floats, floatfix, twocolumn, nofootinbib, amssymb,
               longbibliography]{revtex4-1}

\usepackage{graphicx, bm, hyperref, mathrsfs}
\DeclareSymbolFontAlphabet{\mathrsfs}{rsfs}

\newcommand{\tfrac}[2]{\textstyle \frac{#1}{#2}}
\newcommand{\third}{{\textstyle\frac{1}{3}}}
\newcommand{\half}{{\textstyle\frac{1}{2}}}

\newcommand{\sixth}{{\textstyle\frac{1}{6}}}
\newcommand{\eqref}[1]{(\ref{#1})}

\newcommand{\scri}{\mathrsfs{I}^+\,}
\newcommand{\four}{\,{}^{(4)}}
\newcommand{\three}{\,{}^{(3)}}
\newcommand{\tN}{\tilde N}
\newcommand{\tr}{\mathrm{tr} \,}
\newcommand{\const}{\mathrm{const}}

\begin{document} 
 
\title{Formation and decay of Einstein-Yang-Mills black holes}

\author{Oliver Rinne}
\email[Electronic address: ]{oliver.rinne@aei.mpg.de}
\affiliation{Max Planck Institute for Gravitational Physics 
  (Albert Einstein Institute), Am M\"uhlenberg 1, 14476 Potsdam, Germany}
\affiliation{Department of Mathematics and Computer Science, Freie Universit\"at
 Berlin, Arnimallee 2–6, 14195 Berlin, Germany}

\date{December 30, 2014}

\begin{abstract}
  We study various aspects of black holes and gravitational collapse in 
  Einstein-Yang-Mills theory under the assumption of spherical symmetry.
  Numerical evolution on hyperboloidal surfaces extending to future null 
  infinity is used.
  We begin by constructing colored and Reissner-Nordstr\"om black holes
  on surfaces of constant mean curvature and analyze their perturbations.
  These linearly perturbed black holes are then evolved into the
  nonlinear regime and the masses of the final Schwarzschild
  black holes are computed as a function of the initial horizon radius.
  We compare with an information-theoretic bound on the lifetime of unstable 
  hairy black holes derived by Hod.
  Finally we study critical phenomena in gravitational collapse
  at the threshold between different Yang-Mills vacuum states of the final 
  Schwarzschild black holes, where the $n=1$ colored black hole forms the
  critical solution.
  The work of Choptuik {\it et al.}~[Phys.~Rev.~D 60, 124011 (1999)] 
  is extended by using a family of initial 
  data that includes another region in parameter space where the colored black 
  hole with the opposite sign of the Yang-Mills potential forms the 
  critical solution.
  We investigate the boundary between the two regions and discover that the
  Reissner-Nordstr\"om solution appears as a new approximate
  codimension-two attractor.
\end{abstract}

\maketitle


\section{Introduction}

The Einstein-Yang-Mills equations have proven to be a particularly versatile
model for studying nonlinear phenomena in gravitational collapse.
Whereas for the Einstein-Maxwell equations, a version of Birkhoff's theorem
implies that any spherically symmetric solution must be static, this does
not apply to Yang-Mills fields. 
Hence one can study interesting dynamical behavior in spherical symmetry 
even with modest computational resources.

What makes the Einstein-Yang-Mills equations particularly interesting is the
existence of nontrivial static solutions, which play a dynamical role in
gravitational collapse.
The Bartnik-McKinnon solitons \cite{Bartnik1988} are a discrete family of 
regular static solutions.
There are also black hole solutions, the so-called colored black holes
\cite{Bizon1990,Volkov1990}.
They are \emph{hairy} black holes in the sense that there is Yang-Mills
field outside the horizon but the solutions do not carry a global charge.
This is in contrast to Maxwell and Klein-Gordon fields, where
black hole uniqueness theorems imply that the solutions are characterized
solely by their mass, angular momentum and electric/magnetic charge
evaluated at infinity \cite{HeuslerLRR}.
The Einstein-Yang-Mills equations also admit the  
Reissner-Nordstr\"om black hole as an embedded Abelian 
solution \cite{Yasskin1975}; this solution does carry a global (magnetic) 
charge.
We remark that although we restrict ourselves to spherical symmetry in this
paper, more general axisymmetric regular and black hole solutions to the
Einstein-Yang-Mills equations have been found \cite{Kleihaus1997,Kleihaus1997a}.

There are interesting relations between these spherically symmetric static
solutions.
The colored black holes have the horizon radius $R_h$ as a free parameter.
As $R_h \rightarrow 0$, the solutions approach the Bartnik-McKinnon soliton.
In fact, in \cite{Ashtekar2001} the authors suggested a model of the colored
black hole as a bound state of a bare (Schwarzschild) black hole and
the Bartnik-McKinnon soliton.
And as one sends the index $n$ labeling the member of the discrete family of
colored black holes to infinity, the solutions approach the
Reissner-Nordstr\"om black hole.

All the above static solutions are unstable \cite{Straumann1990a,Straumann1990,
Bizon1991a,Breitenlohner1992,Breitenlohner1995,KahlMA}. 
Within the magnetic ansatz for the Yang-Mills field considered in this paper
(Sec.~\ref{s:EYMeqns}), the $n$th soliton and colored black hole have $n$ 
unstable modes \cite{Lavrelashvili1995,Volkov1995}.
The Reissner-Nordstr\"om solution has an infinite number of unstable modes.
This may come as a surprise because this solution is certainly stable in
Einstein-Maxwell theory.
An explanation in the isolated horizon framework was given 
in \cite{Ashtekar2001}:
the properly defined horizon mass in this framework is \emph{smaller} than the
ADM mass in Einstein-Yang-Mills theory, while the two are the same in
Einstein-Maxwell theory.
This mass difference in the Einstein-Yang-Mills case accounts for the 
available energy that can be radiated away to infinity.

The relevance of these static solutions to the Einstein-Yang-Mills 
equations in the context of the present paper lies in their role as critical 
solutions in gravitational collapse.
Here one starts with a one-parameter family of regular initial data;
depending on the value of the parameter, the outcome of the evolution is
different.
First discovered for the spherically symmetric Einstein-massless scalar field
system by Choptuik \cite{Choptuik1993}, critical phenomena in gravitational
collapse have since been found for a plethora of matter models coupled
to the Einstein equations (see \cite{GundlachLRR} for a review article).
The Einstein-Yang-Mills system is particularly rich in this sense as it
admits both types of critical behavior, as well as a new third type,
depending on the initial data chosen.
Type I and II concern the threshold between dispersal of the field to 
flat spacetime and black hole formation.
In Type I critical collapse, black hole formation turns on at a finite mass, 
and the critical solution is static; in this case it is the Bartnik-McKinnon 
soliton \cite{Choptuik1996}.
In Type II critical collapse, one can make infinitesimally small black holes,
and the critical solution is discretely 
self-similar \cite{Choptuik1996,Gundlach1997}.
A new ``Type III'' critical phenomenon was discovered in \cite{Choptuik1999}.
Here the endstates on \emph{both} sides of the threshold are black holes,
but the Yang-Mills field is in different vacuum states.
It is a curious property of the magnetic ansatz for the Yang-Mills field
in spherical symmetry that there are \emph{two} solutions for the 
connection (or vector potential), namely $w=\pm 1$, that both give rise to 
a vanishing field strength tensor.
At the threshold between black hole formation with $w=\pm 1$,
the $n=1$ colored black hole was found to be the critical 
solution \cite{Choptuik1999}.
Across the threshold there is a \emph{mass gap} between the final Schwarzschild
black holes.

This paper is mainly concerned with ``Type III'' critical collapse.
We have studied this phenomenon for a wide range of parameters; in particular,
we address the question how the mass gap depends on the horizon radius
of the colored black hole critical solution.
Motivated by work on critical collapse in the five-dimensional vacuum 
Einstein equations \cite{Bizon2006}, 
where the authors exploited discrete symmetries to
find a new critical solution with \emph{two} unstable modes by tuning
two parameters in their initial data, we were led to the following question.
The Einstein-Yang-Mills equations are invariant under a sign flip of the
potential, $w \rightarrow -w$.
So each colored black hole solution has a ``dual'' with the opposite sign.
One would expect that this dual solution can also be an attractor in
critical collapse, for different initial data.
Is there a single smooth two-parameter family of initial data that connects 
both regimes?
And if so, what happens at the boundary between them?

Studying these questions requires a robust and efficient numerical code
that is able to carry out accurate long-term evolutions.
A formulation of the Einstein equations on hyperboloidal hypersurfaces 
developed with Vincent Moncrief \cite{Moncrief2009} has proven to be 
extremely useful.
The standard approach to numerical relativity is based on evolution
on Cauchy hypersurfaces truncated at a finite distance, where boundary 
conditions must be imposed.
The hyperboloidal approach avoids any inaccuracies resulting from imperfect 
boundary conditions, as the hypersurfaces
extend all the way to future null infinity $\scri$, where all the 
characteristics leave the computational domain and no boundary conditions need 
to be imposed (or they are determined uniquely by regularity considerations).
In principle we obtain access to the entire future of the initial 
hyperboloidal surface.
Furthermore, the constant mean curvature surfaces we use extend smoothly to the
interior of black hole horizons so that we are able to study gravitational
collapse.

In \cite{Rinne2010} this approach was first implemented for the vacuum
axisymmetric Einstein equations, achieving long-term stable evolutions of
a perturbed Schwarzschild black hole.
In \cite{Rinne2013} we included matter sources and studied power-law
tails of massless scalar and Yang-Mills fields in spherical symmetry.
The present paper uses this latter implementation to study critical phenomena
in gravitational collapse and associated properties of Einstein-Yang-Mills 
black holes.

We would like to mention a few other numerical studies (in addition to those
on critical collapse mentioned above) relevant to the topic of this paper.
In \cite{Puerrer2009}, the authors studied power-law tails for the 
Einstein-Yang-Mills system.
They included $\scri$ in the computational domain by employing Bondi 
coordinates.
Those have the disadvantage that they cannot penetrate black hole
horizons, and hence are unsuitable for gravitational collapse.
In \cite{Zenginoglu2008b}, tails of Yang-Mills fields were computed using
hyperboloidal evolution as in our case, however on a fixed
Schwarzschild background spacetime.
A similar approach was used in \cite{Bizon2010} to study interesting
saddle-point behavior of Yang-Mills fields on a Schwarzschild background.
The test field admits static solutions with one unstable mode, which act as
unstable attractors between basins of attraction belonging to the two different
vacuum states of the Yang-Mills field. 
This is reminiscent of the behavior seen in colored black hole critical 
collapse.

This paper is organized as follows.
In Sec.~\ref{s:1} we briefly review our formulation of the Einstein-Yang-Mills
equations and numerical method.
In Sec.~\ref{s:2} we compute the relevant static solutions on the constant
mean curvature surfaces we use.
We also analyze their perturbations, focusing particularly on the 
Reissner-Nordstr\"om solution.
The eigenvalues are compared with an information-theoretic bound derived by 
Hod \cite{Hod2008}.
In Sec.~\ref{s:3} we turn to nonlinear numerical evolutions.
First we evolve linearly perturbed colored and 
Reissner-Nordstr\"om black holes into the nonlinear regime and investigate
how their final masses depend on the initial horizon radius 
(Sec.~\ref{s:pertevoln}). 
These results are useful for the following Sec.~\ref{s:critcollapse},
which is devoted to critical collapse.
We present an extended family of initial data that includes regions where the
critical behavior found in \cite{Choptuik1999} is reproduced, as well
as regions where the colored black hole with the opposite sign appears as 
the critical solution.
We then investigate the behavior at the threshold between the two and 
demonstrate that the Reissner-Nordstr\"om solution is an approximate
codimension-two attractor.
We conclude and discuss some further questions in Sec.~\ref{s:concl}.


\section{Formulation and methods}
\label{s:1}

In this section we briefly describe the formulation of the Einstein-Yang-Mills
equations and the numerical methods we use.
For simplicity we restrict ourselves to spherical symmetry and a purely magnetic
ansatz for the Yang-Mills field here.
More details and generalizations can be found in \cite{Rinne2013}.


\subsection{Einstein-Yang-Mills equations}
\label{s:EYMeqns}

The Einstein-Yang-Mills equations derive from the action
\begin{equation}
  S = \int d^4 x \; \mu_{\four g} \left( \frac{1}{2\kappa} \four R 
    - \frac{1}{4} F^{(a)}_{\mu\nu} F^{(a)\mu\nu} \right) ,
\end{equation}
where $\four g_{\mu\nu}$ is the spacetime metric,
$\mu_{\four g}$ its volume element, $\kappa = 8\pi$ in geometric units,
$\four R$ is the scalar curvature, and $F^{(a)}_{\mu\nu}$ is the Yang-Mills field 
strength tensor
\begin{equation}  
  F_{\mu\nu}^{(a)} = \partial_\mu A_\nu^{(a)} - \partial_\nu A_\mu^{(a)}
    + \epsilon^{abc} A_\mu^{(b)} A_\nu^{(c)}.
\end{equation}
(For consistency with the literature we have chosen the coupling constant $g$
in \cite{Rinne2013} to be $g=1$ here.)

We write the spacetime metric as 
\begin{equation}
  \label{e:fourg1}
  \four g_{\mu\nu} = \Omega^{-2} \four \gamma_{\mu\nu},
\end{equation}
where the conformal factor $\Omega \searrow 0$ at $\scri$.
In spherical symmetry we may write the conformal metric in isotropic 
coordinates as 
\begin{equation}
  \label{e:fourg2}
  \four \gamma = -\tN^2 dt^2 + (dr + rX \, dt)^2 + r^2 d\sigma^2
\end{equation}
with $d\sigma^2 = d\theta^2 + \sin^2\theta \, d\phi^2$.
We consider an ADM \cite{Arnowitt1962} decomposition with respect to the time 
coordinate $t$. 
Constant mean curvature (CMC) slicing is used; i.e., the mean curvature of the
$t=\const$ slices is a spacetime constant $K>0$.
The tracefree part of the ADM momentum $\pi^{\tr ij}$ has only one degree of
freedom in spherical symmetry, which we take to be
$\pi := (r^4 \sin\theta)^{-1} \pi^{\tr rr}$.
The gravitational field is thus described by the four variables 
$\Omega, \tN, X$ and $\pi$, which are functions of $t$ and $r$ only.
In the following we use an overdot to denote $t$-derivatives and
a prime to denote $r$-derivatives.

Preserving the isotropic spatial coordinate condition and the CMC slicing 
condition under the time evolution yields
\begin{eqnarray}
  \label{e:isotropic}
  0 &=& r^{-1} X' + \tfrac{3}{2} \tN \pi,\\
  \label{e:cmc}
  0&=&-\Omega^2 \tN'' + 3 \Omega\Omega'\tN' - 2\Omega^2 r^{-1}\tN' 
  - \tfrac{3}{2}\Omega'^2 \tN   \nonumber\\
  &&+ \sixth \tN K^2 + \tfrac{15}{8} \tN \Omega^2 r^4 \pi^2
   + \half \kappa \tN \Omega^4 (\tilde S + 2 \tilde \rho).
\end{eqnarray}
The Einstein equations reduce to the Hamiltonian and momentum constraint,
\begin{eqnarray}
  \label{e:hamcons}
  0 &=& -4 \Omega \Omega'' + 6 \Omega'^2 
  - 8 \Omega r^{-1}\Omega' + \tfrac{3}{2} \Omega^2 r^4 \pi^2 \nonumber\\
  &&- \tfrac{2}{3} K^2 + 2 \kappa \Omega^4 \tilde \rho, \\
  \label{e:momcons}
  0 &=& \Omega (r\pi' + 5 \pi) - 2r\Omega' \pi 
  + \kappa \Omega^3 r^{-1} \tilde J^r.
\end{eqnarray}
The source terms $\tilde \rho$, $\tilde S$ and $\tilde J^r$ in 
\eqref{e:cmc}--\eqref{e:momcons} are components
of the conformally rescaled energy-momentum tensor
$\tilde T_{\mu\nu} = \Omega^{-2} T_{\mu\nu}$ and are defined for Yang-Mills fields
below in \eqref{e:matsrc1} and \eqref{e:matsrc2}.

The Yang-Mills equations are conformally invariant and hence we may define
the fields in terms of the conformal metric $\four \gamma_{\mu\nu}$,
indicated by tildes in the following.
We take the gauge group to be SU(2) and adopt a purely magnetic 
ansatz for the vector potential in temporal gauge,
\begin{equation}
  \label{e:ymansatz}
  \tilde A^{i(a)} = [aij] x^j W(t,r), \quad \tilde A_0^{(a)} = 0,
\end{equation}
where the symbol $[aij]$ is totally antisymmetric with $[123] = 1$.
(We considered a more general ansatz in \cite{Rinne2013}.)
In the literature \cite{Bizon1990,Straumann1990,Choptuik1996,Choptuik1999} 
the vector potential is often written in ``(maximally) Abelian 
gauge'' \cite{Bartnik1997} as 
\begin{equation}
  \label{e:abeliangauge}
  \tilde A = w \tau^\theta d\theta + (\cot\theta \, \tau^r + w \tau^\phi)
  \sin\theta \, d\phi
\end{equation}
where $\tau^a$ are the Pauli matrices. 
The two potentials $W$ and $w$ in \eqref{e:ymansatz} and \eqref{e:abeliangauge}
are related by
\begin{equation}
  \label{e:wvsW}
  \qquad W = \frac{1-w}{r^2}.
\end{equation}
We prefer to use $W$ also for numerical reasons (regularity at $r=0$).
The electric field is defined as
\begin{equation}
  \tilde \mathcal{D}^{i(a)} = \mu_{\four \gamma} \, \tilde F^{(a)0i}
  = [aij] x^j D(t,r).
\end{equation}

The Yang-Mills field equations take the form of a nonlinear wave equation 
for $W$:
\begin{eqnarray}
  \label{e:dtW}
  \dot W &=& r X W' + 2 X W - \tN D,\\
  \dot D &=&(rX D - \tN W')' + 2X D -4 \tN r^{-1} W' \nonumber\\
  \label{e:dtD}
           &&-2W r^{-1} \tN' + \tN W^2(r^2 W - 3).
\end{eqnarray}
The matter source terms in \eqref{e:cmc}--\eqref{e:momcons} are given by
\begin{eqnarray}
  \label{e:matsrc1}
  \tilde \rho = \tilde S &=& 
    \half \big[ 2 r^2 D^2 + 12 W^2 + r^2 W^3 (r^2 W - 4) \nonumber\\
    &&\quad + 2 r W' (rW' + 4 W) \big],\\
  \label{e:matsrc2}
  \tilde J^r &=& 2 r D (r W' + 2W).
\end{eqnarray}
%


\subsection{Numerical methods}

We discretize the equations in space using fourth-order finite differences.
A mapping of the radial coordinate with an adjustable 
parameter \cite{Rinne2013} is used
in order to provide more resolution where it is needed, especially near the
black hole horizon where the fields typically develop steep gradients.
The outermost grid point is placed at $\scri$, which we choose to correspond
to $r=1$.
Typical resolutions used for the simulations in this paper range from
$500$ to $2000$ radial grid points.

Following the method of lines, the evolution equations \eqref{e:dtW} and 
\eqref{e:dtD} are first discretized in space and then integrated forward in 
time using a fourth-order Runge-Kutta method with sixth-order
Kreiss-Oliger dissipation \cite{Kreiss1973}.
At each time step, the ODEs \eqref{e:isotropic}--\eqref{e:momcons} are solved
using a Newton-Raphson method, at each iteration solving the resulting
linear system using a direct band-diagonal solver.

Boundary conditions at the origin $r=0$ (before a black hole forms) follow
from the fact that all the fields $\Omega, \tN, X, \pi, W, D$ are even
functions of $r$.
Once an apparent horizon forms, we place an excision boundary sufficiently far
inside it and remove its interior from the computational domain.
One-sided finite differences are used at this inner boundary.
Since the excision boundary lies inside the black hole, all characteristics 
leave the domain and hence no boundary conditions are required for the 
evolution equations \eqref{e:dtW} and \eqref{e:dtD}.
The outer boundary $\scri$ is an outflow boundary as well and is treated in 
the same fashion.
Boundary conditions for the elliptic equations 
\eqref{e:isotropic}--\eqref{e:momcons} follow from regularity at $\scri$ 
and compatibility with the remaining Einstein evolution equations at the
excision boundary; see \cite{Rinne2013} for details.

In all our evolutions the value of the mean curvature is taken to be $K=1/2$.

The code has been written in and the figures produced with 
\texttt{Python}, making use of the \texttt{NumPy}, \texttt{SciPy} and
\texttt{matplotlib} extensions
(for an excellent recent introduction see \cite{Stewart2014}).


\section{Static solutions and their linear perturbations}
\label{s:2}

In order to evolve perturbed Yang-Mills black holes and to analyze 
their role in dynamical collapse evolutions, we need first to construct 
these solutions.
So far these solutions have been considered on maximal slices $K=0$ approaching
spacelike infinity (Sec.~\ref{s:maxsoln}), 
whereas we use CMC slices $K>0$ approaching future null infinity.
We develop a general procedure for transforming a static solution on a maximal
slice to a CMC slice in Sec.~\ref{s:cmcsoln}.
Finally we find the unstable eigenmodes of the solutions in 
Sec.~\ref{s:perttheory}.


\subsection{Solutions on a maximal slice}
\label{s:maxsoln}

Static spherically symmetric solutions to the Einstein-Yang-Mills equations 
are most commonly constructed in polar-areal coordinates, in which the 
spacetime metric takes the form
\begin{eqnarray}
  \label{e:statmetricmax}
  \four g &=& -\left( 1 - \frac{2m}{R}\right) e^{-2\delta} dT^2 
  + \left( 1 - \frac{2m}{R}\right)^{-1} dR^2 \nonumber\\
  &&+ R^2 d\sigma^2, 
\end{eqnarray}
where $m$ and $\delta$ are functions of the areal radius $R$ only.

An example in closed form is the familiar Schwarzschild solution, 
a vacuum solution to the Einstein equations with
\begin{equation}
  m = M = \const, \quad \delta = 0, \quad w = \pm 1, \quad D = 0, 
\end{equation}
$M$ being the black hole mass, $M=0$ corresponding to Minkowski spacetime.
Note there are \emph{two} different solutions for the Yang-Mills potential, 
$w = \pm 1$, that both give rise to vacuum $\tilde F^{(a)}_{\mu\nu} = 0$.
This plays a crucial role in critical collapse where the end state can be
a Schwarzschild black hole with the Yang-Mills field in either of its two vacua.

Another closed-form static solution to the Einstein-Yang-Mills 
equations relevant to the present paper is the Reissner-Nordstr\"om 
solution \cite{Yasskin1975} (with unit magnetic charge)
\begin{equation}
  m = M - \frac{1}{2R}, \quad \delta = 0, \quad w = 0, \quad D = 0.
\end{equation}

Further spherically symmetric, static, purely magnetic solutions to the 
Einstein-Yang-Mills equations have been found:
a discrete family of regular solutions known as Bartnik-McKinnon 
solitons \cite{Bartnik1988} and a discrete family of colored black 
holes \cite{Bizon1990,Volkov1990}.
These solutions are not known in closed form but can be constructed numerically
by solving the static Einstein-Yang-Mills equations, which read
\begin{eqnarray}
  \label{e:eymstatic1}
  0 &=& R^2 \left( 1 - \frac{2m}{R}\right) w_{,RR}  \nonumber\\
  &&+ \left[ 2m - \frac{(1-w^2)^2}{R} \right] w_{,R}
  + (1-w^2)w,\\
  \label{e:eymstatic2}
  m_{,R} &=& \left( 1 - \frac{2m}{R}\right) 
    \left( w_{,R} \right)^2 + \frac{(1-w^2)^2}{2R^2},\\
  \label{e:eymstatic3}
  \delta_{,R} &=&  - \frac{2}{R} (w_{,R})^2.
\end{eqnarray}

We proceed as in \cite{Bizon1990}.
Expanding the solution about the black hole event horizon at $R = R_h$, 
where $R_h = 2m(R_h)$, one finds that given a value of $R_h$, there is one
undetermined parameter at the horizon, namely $b := w(R_h)$.
We integrate the pair of ODEs \eqref{e:eymstatic1} and 
\eqref{e:eymstatic2} numerically from the horizon to some large value of $R$
and, using the shooting method, search for a value of $b>0$ such that
$w \rightarrow -1$ as $R \rightarrow \infty$.
Here we are only interested in the solution with one zero of $w$, the
$n=1$ colored black hole. 
Once the solution for $w$ and $m$ is found, \eqref{e:eymstatic3} is solved 
for $\delta$.
We choose the initial value of $\delta$ at the horizon to be 
$\delta_h = 0$; later a constant is subtracted so that
$\delta \rightarrow 0$ as $R \rightarrow \infty$, ensuring that the metric
\eqref{e:statmetricmax} approaches the flat metric in the standard coordinates
as $R \rightarrow \infty$.

Note that because \eqref{e:eymstatic1}--\eqref{e:eymstatic3} are invariant
under $w\rightarrow -w$, each static solution has a related solution
with the opposite sign of $w$. 
This will play an important role in Sec.~\ref{s:critcollapse}.


\subsection{Transforming to a CMC slice}
\label{s:cmcsoln}

Given the metric \eqref{e:statmetricmax} on the maximal slice, we now show
how to transform it to a CMC slice and a conformally compactified radial
coordinate.
This generalizes the method used in \cite{Malec2003} to find the 
Schwarzschild solution on CMC slices to arbitrary static spherically symmetric 
spacetimes.

First we introduce a new time coordinate
\begin{equation}
  t = T - h(R),
\end{equation}
where $h(R)$ is the height function. 
The mean curvature of a $t=\const$ slice is found to be
\begin{equation}
  \label{e:meancurvature}
  K = e^\delta R^{-2} \frac{d}{dR} \left\{ \frac{R^2 e^{-\delta} h_{,R} 
    \left(1 - \frac{2m}{R} \right)}{\left[ \left(1-\frac{2m}{R}\right)^{-1}
    e^{2\delta} - (h_{,R})^2 \left(1-\frac{2m}{R} \right)\right]^{1/2}}\right\} .
\end{equation}
We require this to be a constant $K>0$.
Let us introduce the function
\begin{equation}
  \label{e:Delta}
  \Delta(R) := \frac{3}{R^3} \left( \int_{R_h}^R \bar R^2 e^{-\delta(\bar R)} d\bar R
  + C_h \right),
\end{equation}
where $C_h$ is an integration constant discussed below, 
and $R_h$ is the areal radius of the horizon.
Eq.~\eqref{e:meancurvature} can now be solved for $h_{,R}$ to obtain
\begin{equation}
  \label{e:hR}
  h_{,R} = \frac{\third K \Delta e^{2\delta}}{\left(1-\frac{2m}{R}\right) f},
\end{equation}
where we have defined
\begin{equation}
  f := \left[ \frac{1}{R^2} \left(1-\frac{2m}{R}\right) + e^{2\delta} 
    \left( \third K \Delta \right)^2 \right]^{1/2}.
\end{equation}
Using \eqref{e:hR} we can write the metric with respect to the new $t$ coordinate as
\begin{equation}
  \four g = -N^2 dt^2 + \three g_{RR} (dR + N^R dt)^2 + R^2 d\sigma^2,
\end{equation}
where
\begin{equation}
  N = e^{-\delta} R f, \quad N^R = -\third K \Delta R^2 f, \quad 
  \three g_{RR} = (Rf)^{-2}.
\end{equation}

Next we transform to conformal radius $r$ defined by
\begin{equation}
  \label{e:confradius}
  \frac{dr}{dR} = \frac{r [\three g_{RR}]^{1/2}}{R} = \frac{r}{R^2 f}.
\end{equation}
Thus we arrive at the line element \eqref{e:fourg1}--\eqref{e:fourg2} with
\begin{equation}
  \Omega = \frac{r}{R}, \quad \tN = r e^{-\delta} f, \quad X = -\third K \Delta.
\end{equation}
We also compute
\begin{equation}
  \pi = \tfrac{2}{3} K R r^{-3} (1 - e^\delta \Delta),
\end{equation}
and from $\dot W = 0$ in \eqref{e:dtW} we find the electric field
\begin{equation}
  D = \third K \Delta e^\delta r^{-3} R^2 w_{,R}.
\end{equation}
This completes the computation of all the evolved fields.

Numerically, we need to integrate \eqref{e:Delta} and \eqref{e:confradius}
to find $\Delta(R)$ and $r(R)$. 
The conformal radius $r$ may be multiplied by an overall constant so that $r=1$ 
at $R = \infty$ corresponding to $\scri$.
The function $r(R)$ is inverted (by interpolation) to find $R(r)$ on the 
numerical grid chosen for the conformal radius $r$, and this then allows us 
to specify all the fields as functions of $r$ on the numerical grid.

Finally we comment on the integration constant $C_h$ in \eqref{e:Delta}.
The expansion of the outgoing radial null rays is computed as
\begin{equation}
  \Theta_+ = \third K \Delta e^\delta + f.
\end{equation}
At the horizon, we have $R_h = 2 m_h$ and thus
\begin{equation}
  \Theta_{+, h} = \third K e^{\delta_h} (\Delta_h  + \vert\Delta_h\vert). 
\end{equation}
Hence in order to have $\Theta_{+, h} = 0$ we need the constant $C_h < 0$
in \eqref{e:Delta}.
In Schwarzschild (and also Reissner-Nordstr\"om) spacetime we have 
$\delta \equiv 0$, and the constant $C$ appearing in the metric on the CMC
slice in \cite{Malec2003} is related to our constant $C_h$ via
\begin{equation}
  C_h = -\frac{C}{K} + \frac{1}{3} R_h^3.
\end{equation}
%


\subsection{Mode analysis}
\label{s:perttheory}

In order to compute the eigenmodes of linear perturbations about static 
solutions of the Einstein-Yang-Mills equations, we prefer to work on the CMC
slice using conformal radius. 
This is because our coordinates smoothly extend to the
black hole interior, unlike the standard Schwarzschild-like coordinates, 
or Kruskal-Szekeres coordinates where the horizon is pushed off to 
negative infinity in the radial coordinate used
(cf.~the debate on smoothness of the modes 
in \cite{Straumann1990,Bizon1991,Bizon1991a}).
Also our computational domain extends all the way out to future null infinity.

Separating the time dependence, we make the general ansatz
\begin{equation}
  \label{e:pertansatz}
  w = \bar w + e^{\lambda t} \delta w (r),
\end{equation}
with $\lambda \in \mathbb{C}$, where $\bar w$ refers to the static background 
solution and $\delta w$ to its linear perturbation. 
We are looking for unstable modes corresponding to positive real eigenvalues 
$\lambda > 0$.
Substituting the ansatz \eqref{e:pertansatz} into the dynamical 
Einstein-Yang-Mills equations (Sec.~\ref{s:EYMeqns}), we find the pulsation 
equation
\begin{equation}
  \label{e:pulseqn}
  c_2 \delta w'' + c_1 \delta w' + c_0 \delta w = 0,
\end{equation}
where as in Sec.~\ref{s:EYMeqns} a prime refers to a derivative with respect to conformal
radius $r$.
The coefficients $c_i$ are given by
\begin{eqnarray}
  \label{e:c2}
  c_2 &=& \tN^2 - r^2 X^2,\\
  c_1 &=& \tN^{-1} \tN' (\tN^2 + r^2 X^2) - 2r^2 XX' - 2rX^2 \nonumber\\
     &&+ 2 \lambda rX,\\
  c_0 &=& -2 r^{-2} \tN^2 + 3 \tN^2 r^{-2} (1-\bar w^2)\nonumber\\
  && + \lambda (X - \tN^{-1} \tN' rX + rX') - \lambda^2,
\end{eqnarray}
where $\tN$ and $X$ are to be evaluated for the background solution.
At $\scri$ the following conditions hold \cite{Rinne2013}:
\begin{equation}
  r\tN' = \tN, \quad rX = -\tN, \quad X' = 0.
\end{equation}
Substituting these into \eqref{e:pulseqn} yields
\begin{equation}
  \label{e:pulsini}
  \delta w' = - \delta w \, \frac{\hat \lambda^2 + 2 r^{-2}}{2\hat \lambda}
\end{equation}
at $\scri$, where we have defined $\hat \lambda := \lambda/\tN_\infty$,
and we choose the lapse at $\scri$ to be $\tN_\infty = K/3$ so that time $t$
agrees with the standard time coordinate in flat spacetime 
asymptotically \cite{Rinne2013}.
The value of $\delta w$ at $\scri$ can be freely chosen, and \eqref{e:pulsini}
provides us with initial data for the pulsation equation \eqref{e:pulseqn} 
at $\scri$, which we integrate inwards toward the horizon. 
There the solution $\delta w$ must be finite, and this condition determines the
eigenvalues $\lambda$.
Once an eigenvalue has been found, \eqref{e:pulseqn} can be integrated 
to the interior of the horizon as well.

For the $n=1$ colored black hole, there is precisely one unstable mode, 
which was first computed in \cite{Straumann1990}.
A plot of the eigenvalue as a function of the horizon radius can be found 
in \cite{Bizon2000}.
Our results are in good agreement with this.
 
\begin{figure}
\centerline{\includegraphics[width=.475\textwidth]{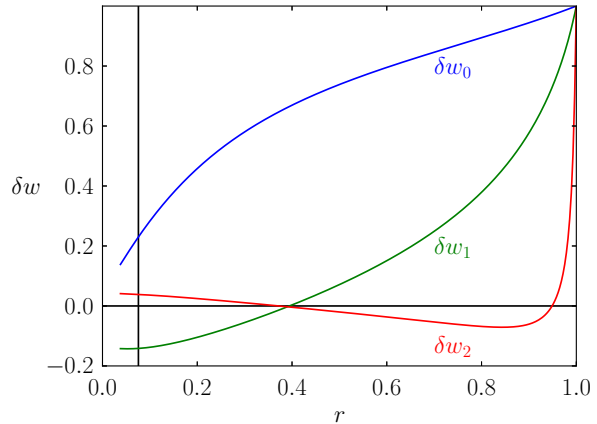}}
\caption{\label{f:RN_modes}
  The first three unstable eigenmodes of the Reissner-Nordstr\"om solution
  as a function of conformal radius $r$. 
  The horizon radius $R_h = 2.57$ and slicing constant $C_h = -1.84$
  in \eqref{e:Delta} have been chosen to agree with the 
  attractor observed in the evolutions of Sec.~\ref{s:critcollapse}.
  The modes have been normalized so that $\delta w = 1$ at $\scri$.
  The location of the horizon is indicated by the vertical line.
}
\end{figure}

\begin{figure}
\centerline{\includegraphics[width=.475\textwidth]{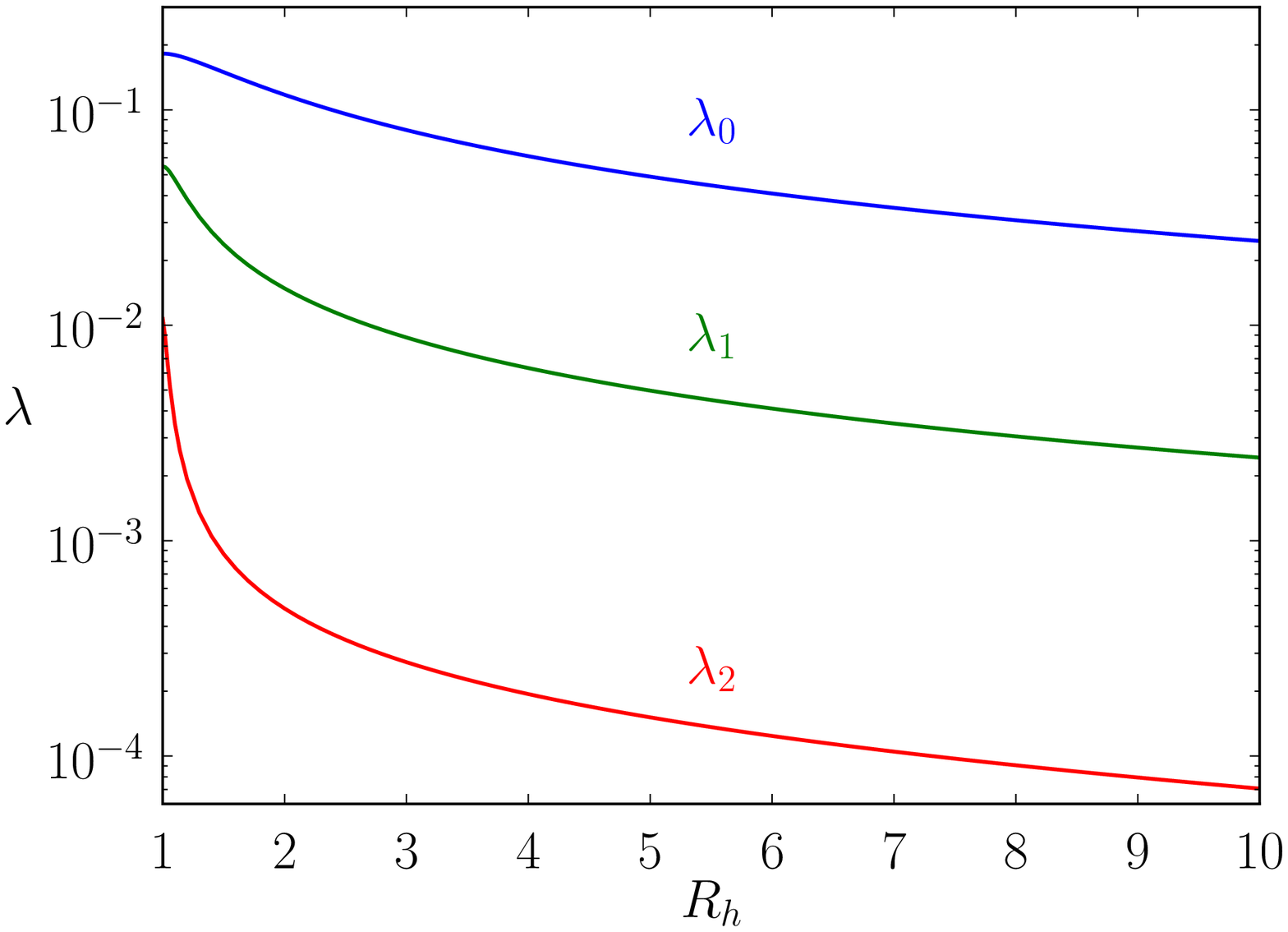}}
\centerline{\includegraphics[width=.475\textwidth]{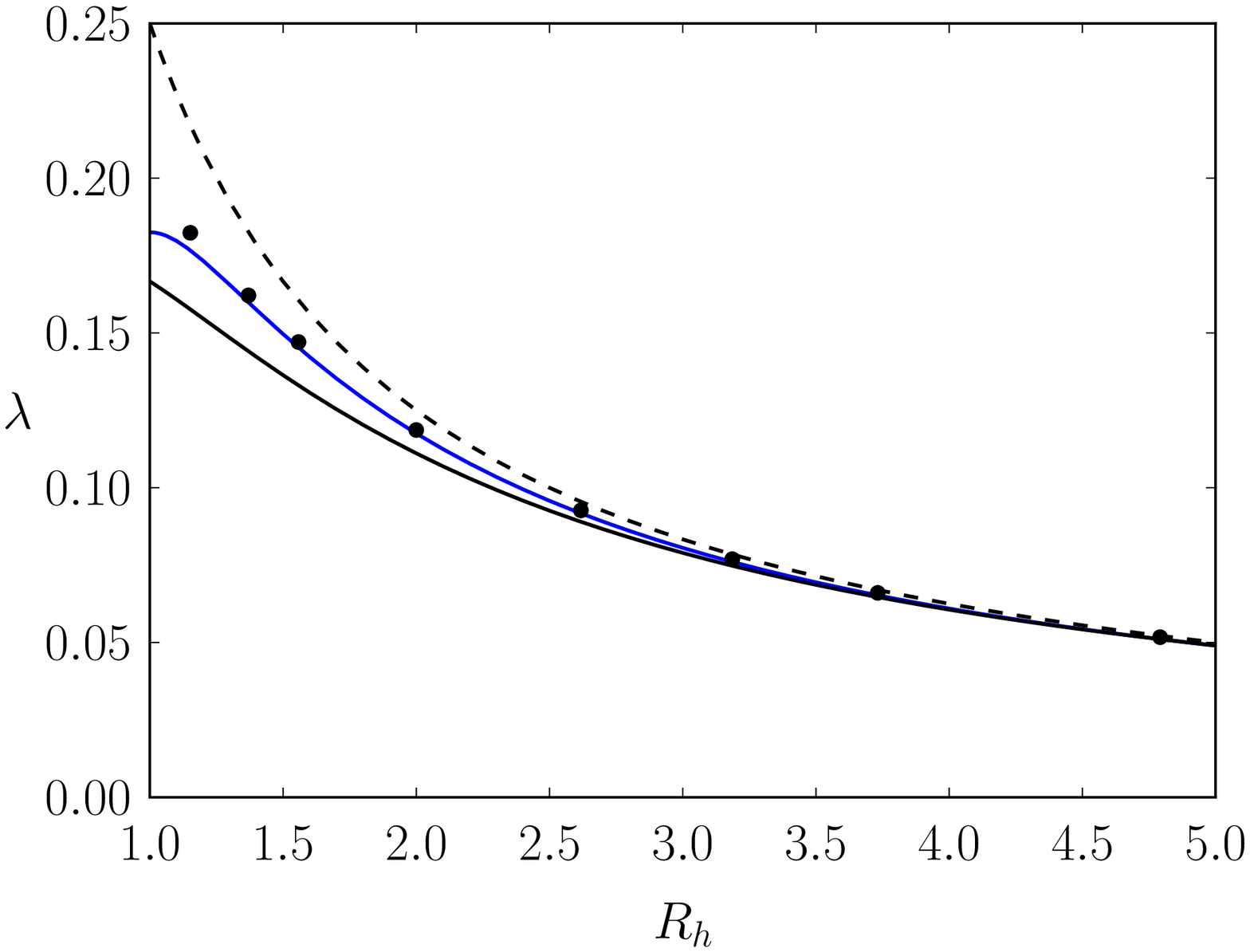}}
\caption{\label{f:RN_EVs}
  Top: the first three unstable eigenvalues of the Reissner-Nordstr\"om solution
  as functions of horizon radius $R_h$.
  Bottom: comparison of the dominant eigenvalue (middle curve, solid blue) 
  with Hod's bound \cite{Hod2008} in its weak (upper curve, dashed black) 
  and strong (lower curve, solid black) form,
  where for the latter we take $\Delta \mathcal{E}$ to be the entire mass
  outside the horizon.
  If instead $\Delta \mathcal{E}$ is taken to be the fraction of this mass 
  that falls into the black hole in a nonlinear evolution, the black points
  result.
}
\end{figure}

In the Einstein-Yang-Mills system, the Reissner-Nordstr\"om solution has an 
infinite number of unstable modes \cite{Bizon1991a,Breitenlohner1995,KahlMA}.
The $n$th mode has $n$ zeros, $n=0, 1, 2, \ldots$
The first three modes are plotted in Fig.~\ref{f:RN_modes}
for one value of the horizon radius.
Note how they are perfectly smooth at the horizon in our coordinates.
Fig.~\ref{f:RN_EVs} shows the corresponding eigenvalues as functions of
horizon radius. 
 
It is interesting to compare the eigenvalues of the
Reissner-Nordstr\"om solution with a bound on the lifetime of
unstable hairy black holes proposed by Hod \cite{Hod2008}, based on arguments 
from quantum information theory.
This bound was shown in \cite{Hod2008} to be satisfied for the $n=1$ colored 
black hole but the situation for the Reissner-Nordstr\"om solution remained
inconclusive.
The bound implies that the eigenvalues should be bounded by
\begin{equation}
  \label{e:hod}
  \lambda \leqslant [4 (R_h + \Delta \mathcal{E})]^{-1} \leqslant (4 R_h)^{-1}.
\end{equation}
We refer to the first bound on the right-hand side as the strong bound and
the second as the weak bound.
In the strong bound, $\Delta \mathcal{E}$ is the mass that is swallowed by the
unstable black hole in a nonlinear evolution.
As a first approximation, this was taken in \cite{Hod2008} 
to be the entire mass outside the horizon of the initial black hole,
$\Delta \mathcal{E} = \Delta M_\mathrm{max} = M - \half R_h$.
For the Reissner-Nordstr\"om solution, we have
\begin{equation}
  \label{e:DeltaE}
  M = \frac{R_h^2 + 1}{2 R_h} \, \Rightarrow \, 
  \Delta M_\mathrm{max} = M - \frac{R_h}{2} = \frac{1}{2 R_h}.
\end{equation}

Figure \ref{f:RN_EVs} shows that close to the extremal value $R_h = 1$,
only the weak bound is satisfied, whereas the above version of the strong bound
is violated
(unlike for the colored black holes, where both are satisfied \cite{Hod2008}).
However, $\Delta \mathcal{E}$ should really be taken to be the actual
amount of hair that falls into the black hole; this will be computed in the
following section.


\section{Nonlinear evolutions}
\label{s:3}

In this section we perform nonlinear numerical evolutions of the 
Einstein-Yang-Mills equations.
While we are ultimately interested in the formation (and later decay) 
of Yang-Mills black holes as intermediate attractors in gravitational 
collapse (Sec.~\ref{s:critcollapse}), we begin by studying the final fate 
of linear perturbations of these static solutions (Sec.~\ref{s:pertevoln}).
The reason is that in critical collapse, the situation near the
critical solution can be described in terms of linear perturbations.
As we shall see, the critical solutions relevant to the present study are the
$n=1$ colored black holes and the Reissner-Nordstr\"om solution.
In order to approach these critical solutions, one needs to tune
one (or even two, as in Sec.~\ref{s:critcollapse}) parameters in the initial
data, which is done by a bisection search involving a large number of 
evolutions.
If we are only interested in the behavior after the critical solution is 
approached, it suffices and is much less time-consuming to start with the 
linearly perturbed critical solution and evolve it into the nonlinear regime.


\subsection{Perturbed black holes}
\label{s:pertevoln}

\begin{figure}
\centerline{\includegraphics[width=.475\textwidth]{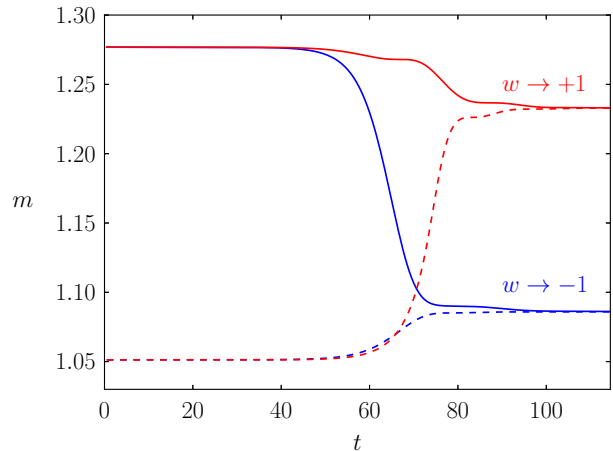}}
\caption{\label{f:CBH_mass_evoln}
  Bondi mass $m_{\scri}$ (solid lines) and apparent horizon mass 
  $m_h = \half R_h$ (dashed lines) as functions of time for two perturbed 
  colored black hole evolutions with opposite sign of the perturbation,
  leading to final vacuum states $w=-1$ (blue) or $w=+1$ (red).
  The initial horizon radius is $R_h = 2.11$ (chosen to agree with the 
  attractor of the evolution with $r_b = 0.7$ in Sec.~\ref{s:critcollapse}).
}
\end{figure}

\begin{figure}
\centerline{\includegraphics[width=.475\textwidth]{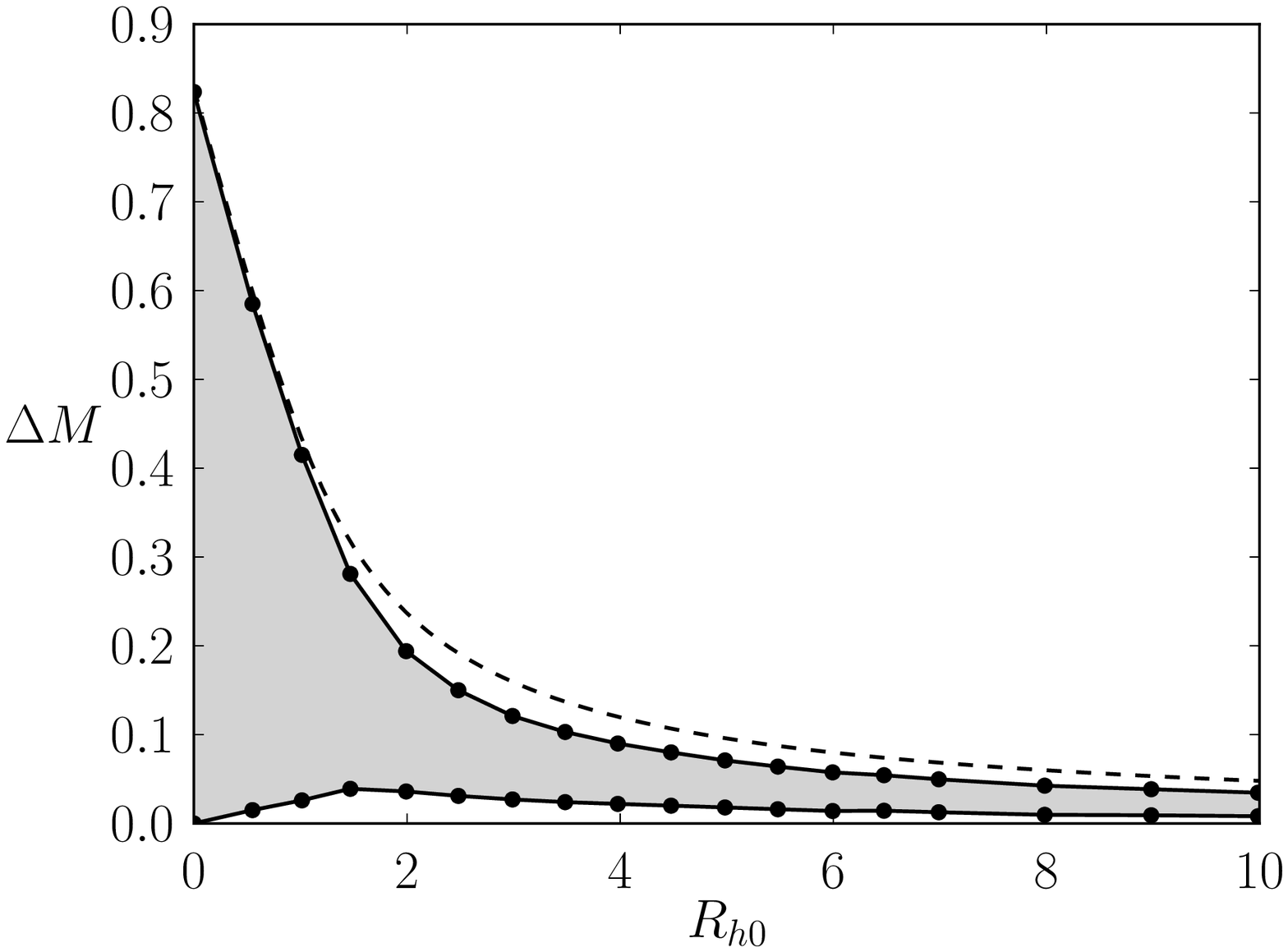}}
\centerline{\includegraphics[width=.475\textwidth]{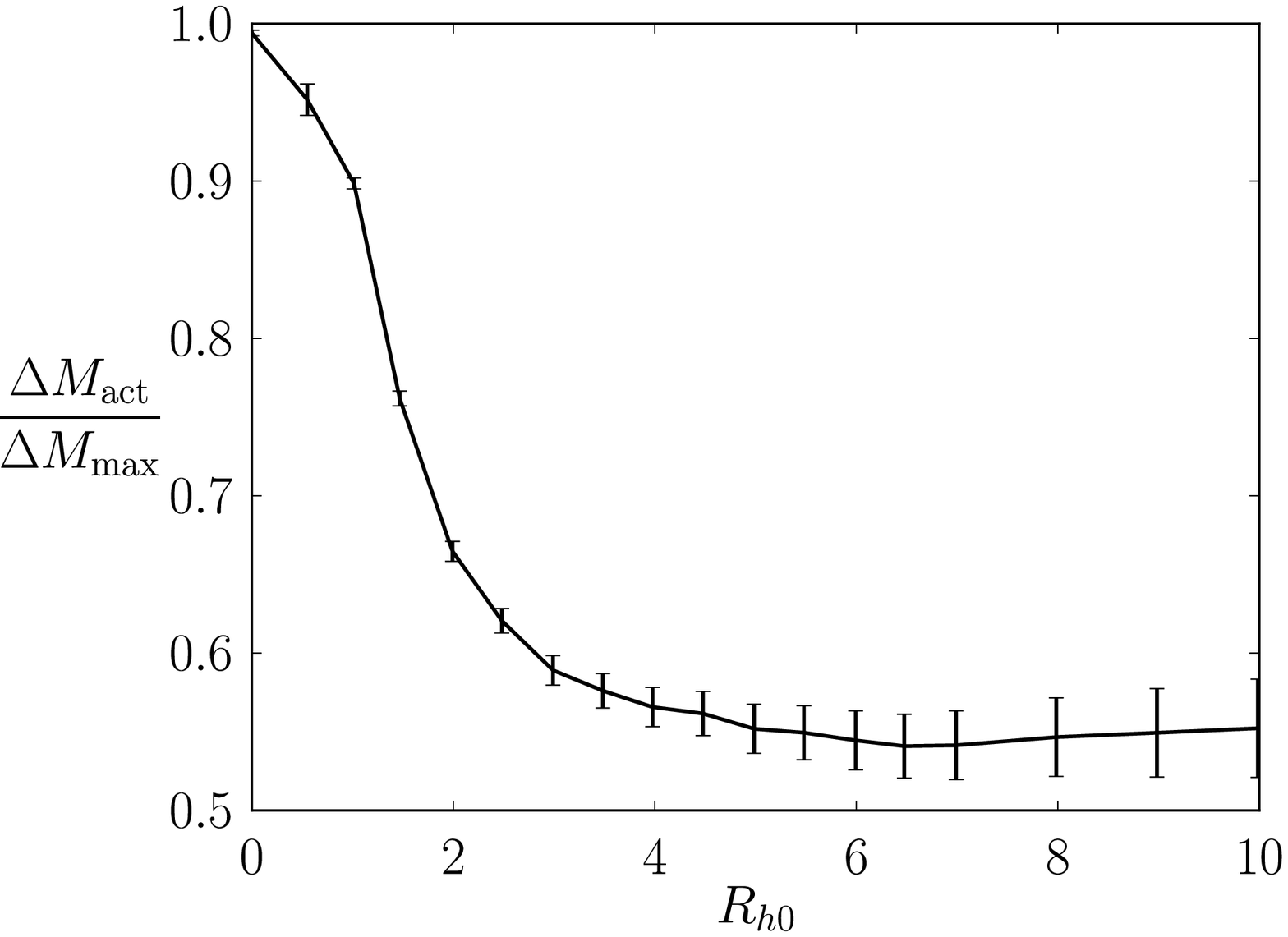}}
\caption{\label{f:CBH_masses}
  Masses of colored black holes and their decay products. 
  In the top panel, the dashed line shows the total mass of hair 
  $\Delta M_\mathrm{max} = M - \half R_{h0}$ 
  outside the horizon radius $R_{h0}$ of the initial colored black hole.
  The upper solid line represents the quantity $M_{f+} - \half R_{h0}$,
  where $M_{f+}$ is the mass of the final Schwarzschild black hole in a
  $w \rightarrow +1$ evolution.
  The lower solid line shows the corresponding $M_{f-} - \half R_h$ for a
  $w \rightarrow -1$ evolution.
  The difference $\Delta M_\mathrm{act} = M_{f+} - M_{f-}$ is the actual mass gap,
  indicated by the shaded region.
  The ratio of the actual and the maximal mass gap is plotted 
  in the lower panel.
  The error bars indicate an estimate of the numerical accuracy.
}
\end{figure}

We begin with a colored black hole solution of a given horizon radius $R_{h0}$
and add to it the unstable mode with a typical amplitude 
$\sim 5 \times 10^{-3}$.
Depending on the sign of the perturbation, the solution evolves to a 
Schwarzschild black hole with the Yang-Mills field either in its $w=+1$ or
$w=-1$ vacuum state. 
Fig.~\ref{f:CBH_mass_evoln} illustrates that in a $w\rightarrow +1$ evolution,
most (but not all) of the Yang-Mills hair falls into the black hole, 
whereas in a $w\rightarrow -1$ evolution, most (but not all) of it escapes
to infinity.
The resulting mass gap is plotted in Fig.~\ref{f:CBH_masses} as a function of
the horizon radius of the initial colored black hole for a few evolutions.
Note that the case $R_h = 0$ corresponds to the regular, horizon-less 
Bartnik-McKinnon soliton, which either disperses to Minkowski spacetime
or collapses to a Schwarzschild black hole swallowing nearly all of 
the soliton's mass.

In Fig.~\ref{f:CBH_masses} we compare the actual mass gap observed in 
the nonlinear evolutions with the \emph{maximal} mass gap
(see Fig.~2 of \cite{Bizon2000}) that would result if either \emph{all} 
of the Yang-Mills hair outside the horizon fell into the black hole or 
\emph{all} of it escaped to infinity.
In \cite{Bizon2000} the authors wondered whether the actual mass gap goes to 
zero at some large but finite $R_h$, so that the line of colored black holes
in a phase space diagram \cite{Choptuik1999} would terminate at a 
finite distance,
``in an amusing similarity to the gas-liquid boundary on phase diagrams
for typical substances''.
While our results cannot exclude this possibility, if the trend continues then
the mass gap will only approach zero asymptotically as $R_h \rightarrow \infty$.
We remark that the numerical evolution is stopped when the horizon mass 
$\half R_h$ and the Bondi mass differ by less than one part in $10^4$, 
so we can only determine the final mass up to that accuracy. 
As a result, the relative mass gap, being the quotient of two small quantities,
has relatively large errors for large values of $R_h$.

\begin{figure}
\centerline{\includegraphics[width=.475\textwidth]{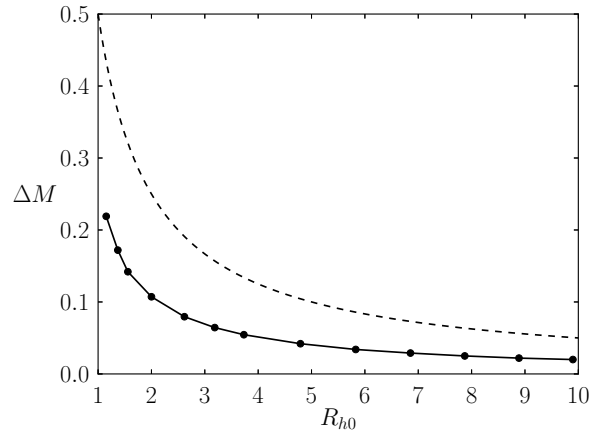}}
\caption{\label{f:RN_masses}
  Masses of Reissner-Nordstr\"om black holes and their decay products. 
  The dashed line shows the mass of the Yang-Mills field 
  outside the horizon radius $R_{h0}$ of the initial Reissner-Nordstr\"om black 
  hole, $\Delta M_\mathrm{max} = M - \half R_{h0}$.
  The solid line represents the quantity $M_f - \half R_{h0}$, where $M_f$ is the
  mass of the final Schwarzschild black hole.
}
\end{figure}

We repeat this calculation for the Reissner-Nordstr\"om solution in
Fig.~\ref{f:RN_masses}, where we add the dominant unstable mode.
Since the background solution has $w=0$ and the Einstein-Yang-Mills equations
are symmetric under $w\rightarrow -w$, the evolutions for both signs of the
perturbation are identical up to a sign, and there is no mass gap.
We have already computed the mass outside the horizon of the 
Reissner-Nordstr\"om solution, $\Delta M_\mathrm{max}$ in \eqref{e:DeltaE}.
The part of this mass that falls into the black hole in a nonlinear evolution
is shown in Fig.~\ref{f:RN_masses}.
It shows remarkably little variation, starting at about $50\%$ of 
$\Delta M_\mathrm{max}$ for $R_{h0} \rightarrow 1$ and settling down to about 
$40\%$ for $R \sim 10$.
Since we cannot evolve the extremal Reissner-Nordstr\"om solution 
($R_h = 1$) using our current numerical implementation, we only consider
horizon radii close to but larger than this limit.

With these results we now return to the comparison of the dominant unstable
eigenvalue of the Reissner-Nordstr\"om solution with the bound derived by 
Hod \cite{Hod2008}, the lower panel of Fig.~\ref{f:RN_EVs}.
If we take $\Delta \mathcal{E}$ in \eqref{e:hod} to be the actual amount of
mass that falls into the black hole during the evolution,
$\Delta \mathcal{E} = M_f - \half R_{h0}$, then the strong version of the
bound is satisfied and saturated remarkably closely.


\subsection{Critical collapse}
\label{s:critcollapse}

In this section we study the formation of black holes from regular initial
data. 
The family of initial data for the Yang-Mills potential $w$ we consider
consists of a kink as in \cite{Choptuik1999} with an additional Gaussian
bump:
\begin{equation}
  \label{e:wini}
  w(0,r) = - \tanh \left(\frac{r - r_k}{\sigma_k}\right) - A_b \exp \left[
  - \frac{(r - r_b)^2}{2 \sigma_b^2} \right].
\end{equation}
From this we form our evolved field $W$, rolling it off with an additional 
Gaussian to ensure regularity at $r=0$:
\begin{equation}
 W(0,r) = \frac{1 - w(0,r)}{r^2} \left[ 1 - \exp \left( -\frac{r^2}{2\sigma_r^2}
   \right) \right].
\end{equation}
The time derivative of $w$ is taken to vanish initially.
The constraint equations are solved for the gravitational field.

In \cite{Choptuik1999} colored black hole critical collapse was observed
in the kink family of initial data (without our additional Gaussian,
$A_b = 0$ in \eqref{e:wini}).
We observe the same phenomenon in regions of our extended family.
Since the Einstein-Yang-Mills equations
are symmetric under $w \rightarrow -w$, it is tempting to search for 
regions in parameter space where the critical solution is the one with the
opposite sign as in the original simulations of \cite{Choptuik1999}.
Indeed such regions exist, and an extensive search in parameter space
led us to discover a two-parameter family that smoothly connects two regions 
with opposite signs of the colored black hole critical solution.
We fix $\sigma_k = \sigma_b = \sigma_r = 0.05$
and $r_k = 0.4$ in \eqref{e:wini} and vary the two parameters
$r_b$ and $A_b$.

\begin{figure}
\centerline{\includegraphics[width=.475\textwidth]{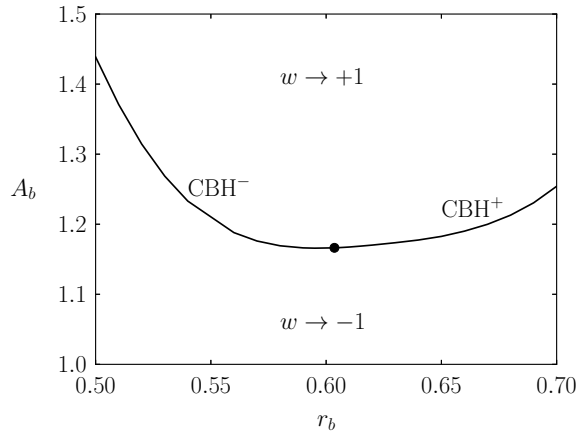}}
\caption{\label{f:parameter_space}
  The critical line in the two-dimensional parameter space.
  For values of $A_b$ above (below) the threshold, the evolution ultimately
  approaches a Schwarzschild black hole with $w\equiv +1$ ($-1$).
  The critical value $r_b = r_b^*$ is indicated by a dot.
  For $r_b < r_b^*$ ($r_b > r_b^*$), the critical solution is a colored black 
  hole (CBH) with $w=-1$ ($w=+1$) at $\scri$.
}
\end{figure}

\begin{figure}
\includegraphics[width=.238\textwidth]{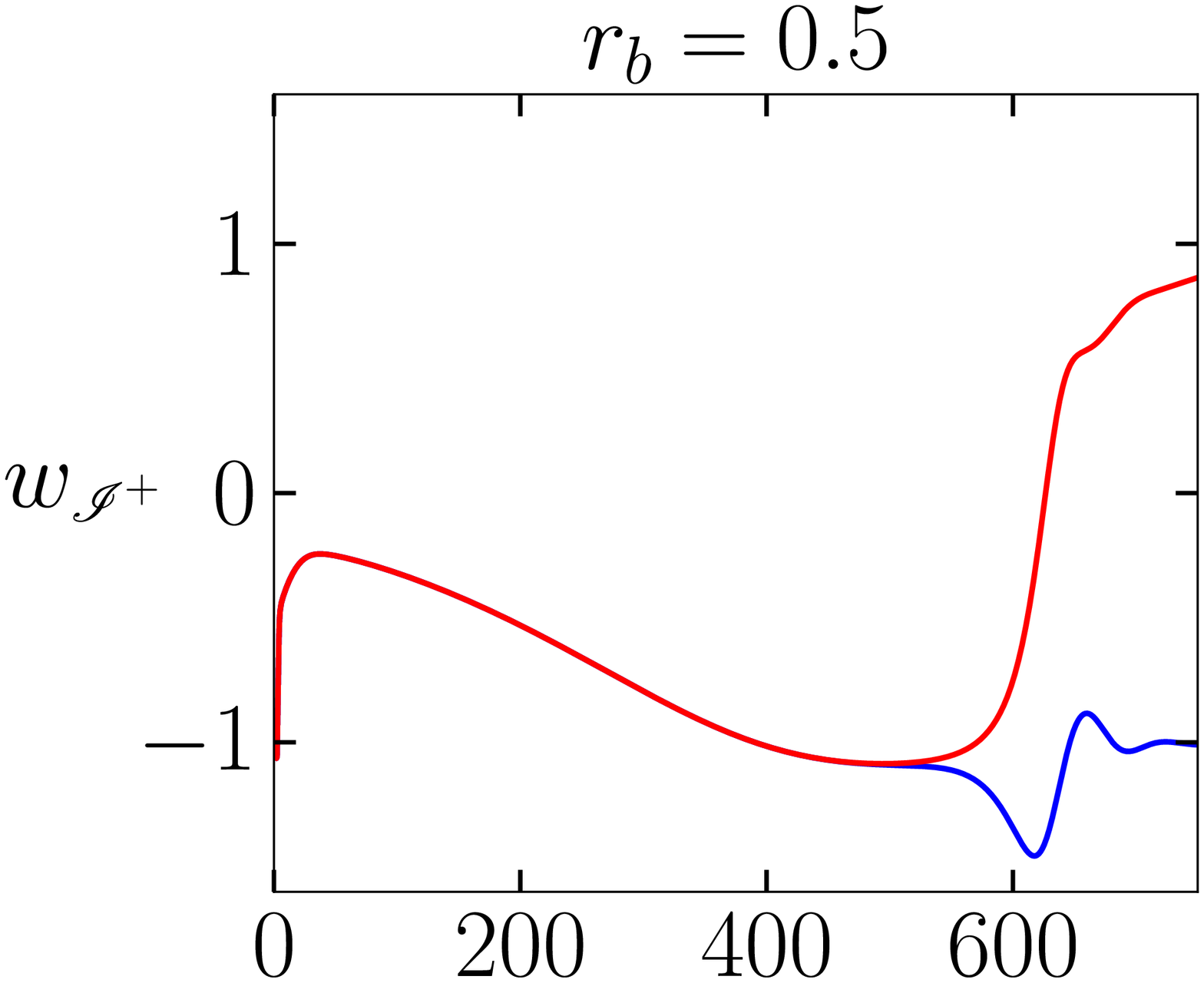}
\includegraphics[width=.238\textwidth]{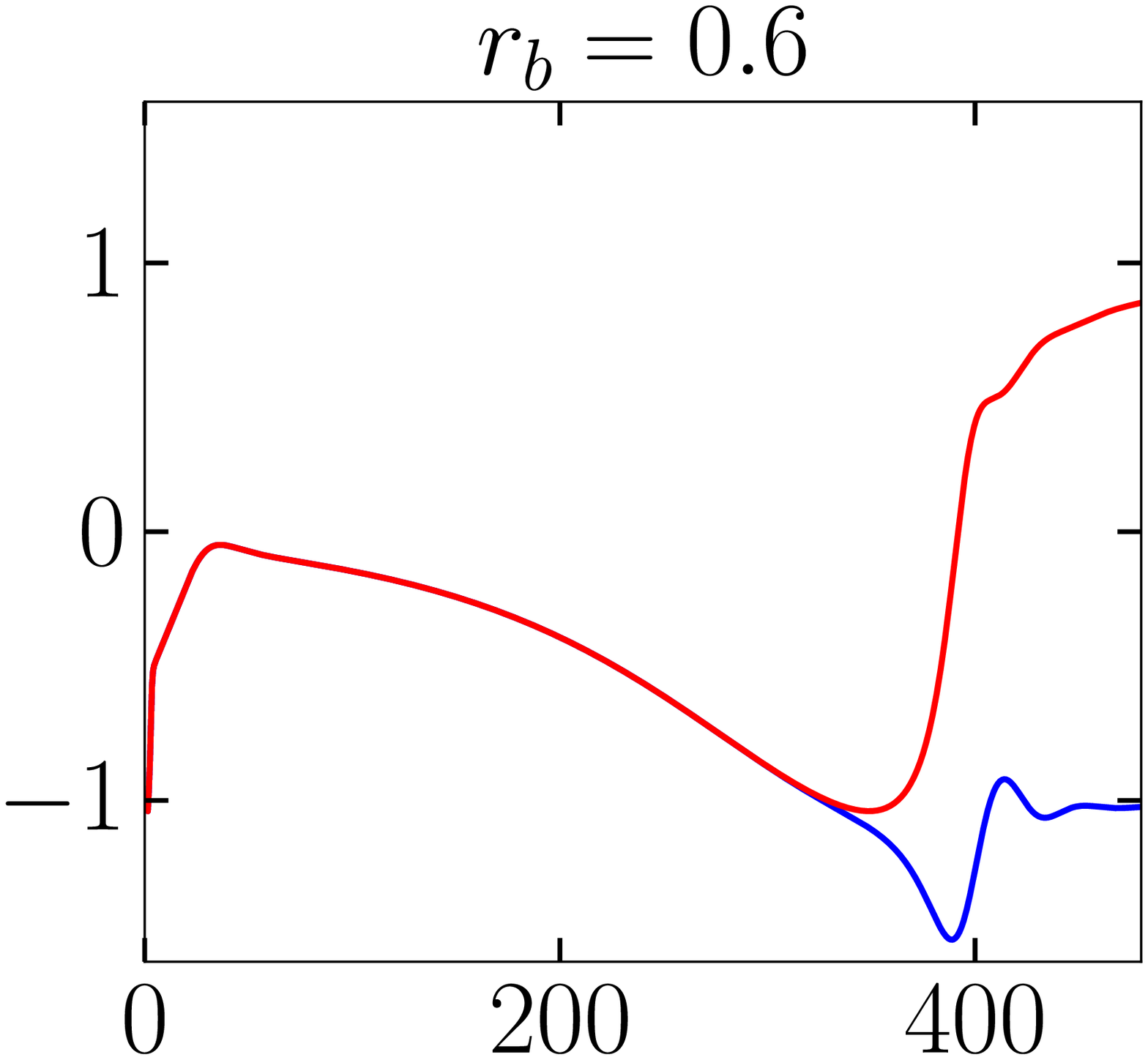}\\
\includegraphics[width=.238\textwidth]{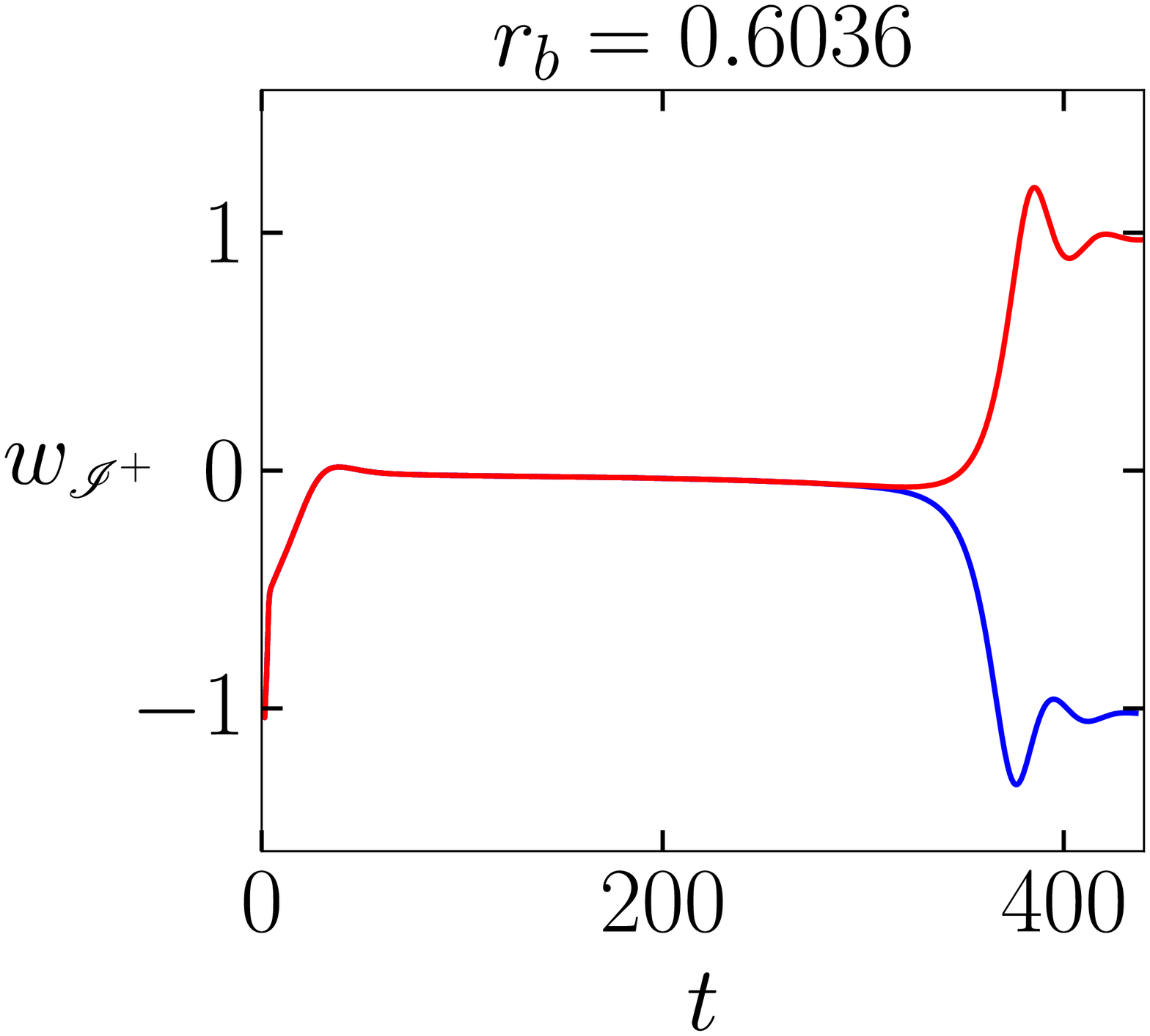}
\includegraphics[width=.238\textwidth]{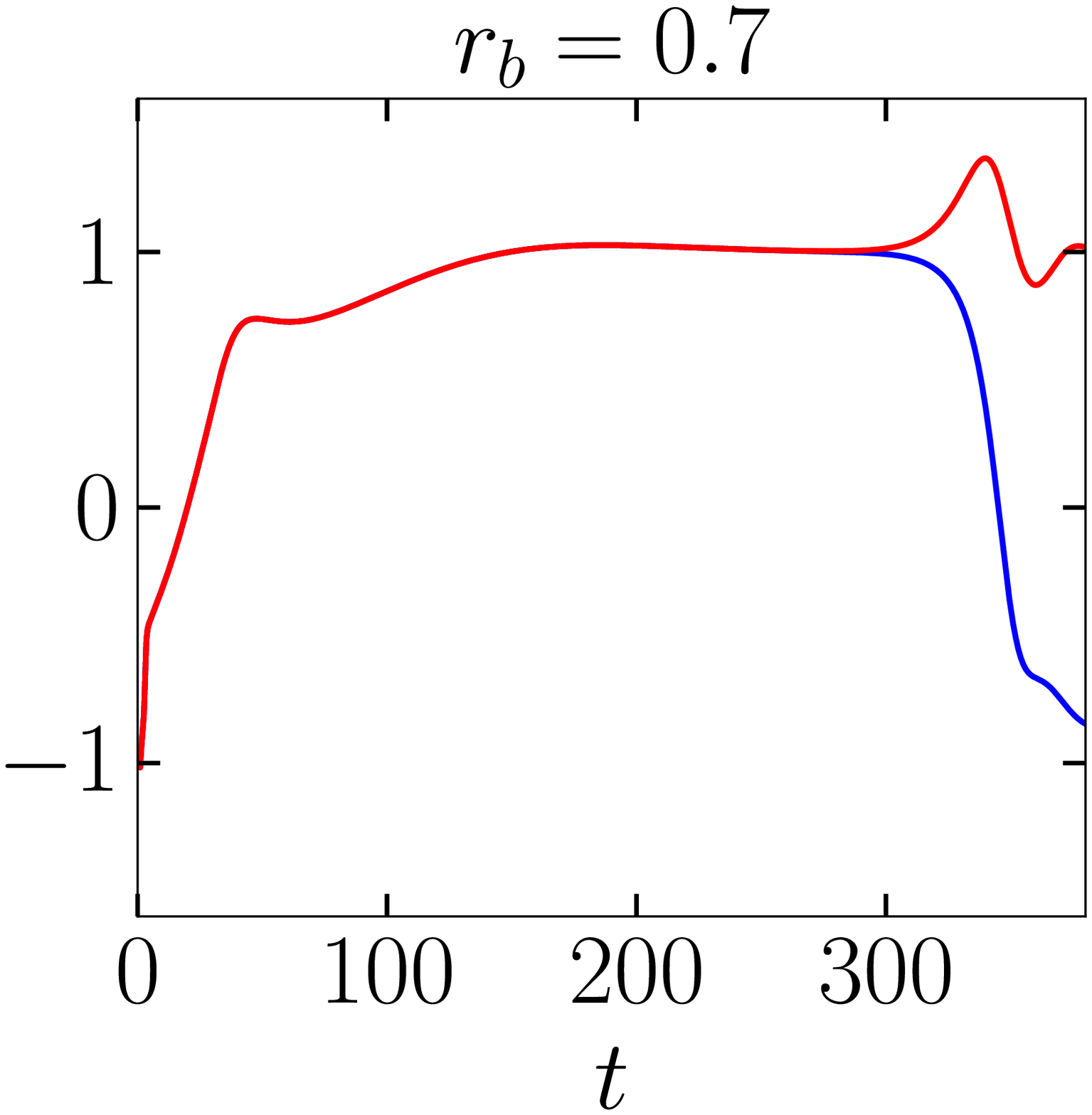}
\caption{\label{f:w_evolns}
  $A_b$-near-critical evolutions of $w$ at future null infinity for a few 
  typical values of $r_b$. 
  The value of $A_b$ for the slightly supercritical (upper curves, red) and
  slightly subcritical (lower curves, blue) evolutions differs by 
  $\approx 10^{-14}$.
}
\end{figure}

For various values of $r_b$, we perform a bisection search for $A_b$
between the two possible outcomes of the evolution, Schwarzschild black holes 
with $w=\pm 1$ (Fig.~\ref{f:parameter_space}).
Let us call the solution that is thus approached $A_b$-critical.
For values of $r_b$ well smaller than some critical value 
$r_b^* \approx 0.6036$, we find that the critical solution is a colored black 
hole with $w=-1$ at $\scri$, as in \cite{Choptuik1999}.
For $r_b > r_b^*$, the colored black hole with the \emph{opposite}
sign of $w$ is approached (Fig.~\ref{f:w_evolns}).

\begin{figure}
\includegraphics[width=.475\textwidth]{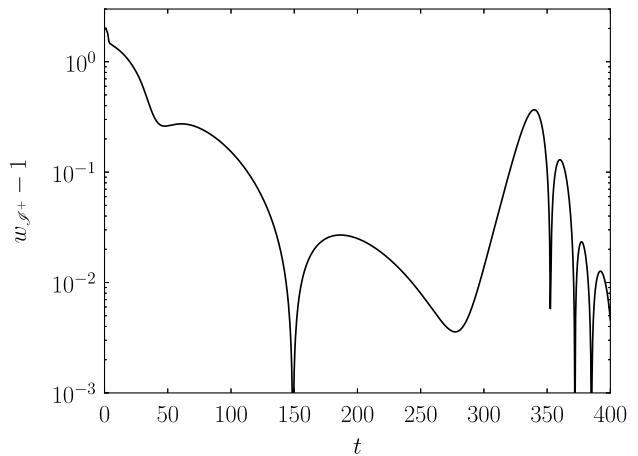}
\caption{\label{f:w_ringdown}
  An ever so slightly $A_b$-supercritical evolution of $w$ at $\scri$ for 
  $r_b = 0.7$. 
}
\end{figure}

The behavior we find at the $A_b$-critical threshold is the same as reported 
in \cite{Choptuik1999}.
Figure \ref{f:w_ringdown} shows $w$ at $\scri$ as a function of time for a 
slightly $A_b$-supercritical evolution with $r_b = 0.7$.
The black hole forms at $t=28$, and the different phases of the evolution are 
clearly visible:
approach to the colored black hole with $w=+1$ at $\scri$, 
exponential instability of this intermediate attractor,
and ringdown to the final Schwarzschild black hole with $w=+1$ at $\scri$.
We only plot $w$ at $\scri$ here, however the approach to the static solutions
can be observed at all radii. 
 
\begin{figure}
\centerline{\includegraphics[width=.475\textwidth]{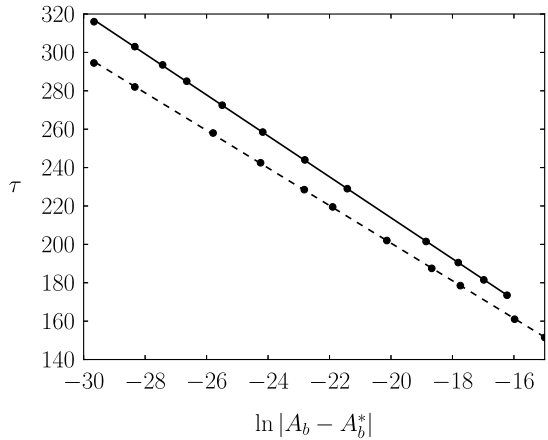}}
\caption{\label{f:Ab_timescaling}
  The time $t=\tau$ when the evolution departs from the critical solution 
  as a function of the logarithm of the parameter distance in $A_b$-critical
  searches. 
  The dashed line is for $r_b = 0.7$, where the critical solution is a colored
  black hole, whereas the solid line is for $r_b = 0.6036 \approx r_b^*$,
  where the critical solution is the Reissner-Nordstr\"om black hole.
  The lines are linear fits to the data points.
}
\end{figure}

The time $t = \tau$ when the evolution departs from the critical solution is 
plotted for this $r_b = 0.7$ evolution as the dashed line in 
Fig.~\ref{f:Ab_timescaling}.
Here we define $\tau$ as the time when the zero of $w$ crosses 
$r=0.5$ in slightly subcritical evolutions.
It exhibits critical scaling as in Type I critical collapse \cite{GundlachLRR},
\begin{equation}
  \tau = \const - \gamma \ln \vert A_b - A_b^* \vert
\end{equation}
with a fitted value $\gamma = 9.79$.
Using the methods of Sec.~\ref{s:perttheory} we compute the eigenvalue of 
the unstable mode of the colored black hole (with $R_h = 2.11$ for this 
evolution) as $\lambda = 0.1020$.
Hence $1/\lambda = 9.80 \approx \gamma$, as expected.
Our results demonstrate the universality of the critical phenomenon discovered
in \cite{Choptuik1999}, as the family of initial data and the critical parameter
we use is different (recall the Gaussian in \eqref{e:wini} was not included
in \cite{Choptuik1999}).

We also observe a mass gap between the slightly $A_b$-subcritical and 
slightly $A_b$-supercritical evolutions, and this agrees well with the mass gap
we obtained in Sec.~\ref{s:pertevoln} by starting off with the perturbed 
colored black hole directly.
(For the case $R_h = 2.11$ considered above, the mass gap between
near-critical evolutions is found to be $\Delta M = 0.1465$; 
the mass gap from the perturbed colored black hole evolutions is 
$\Delta M = 0.1469$.)

The question arises what happens as $r_b$ approaches the critical value
$r_b^*$ that separates the two regions of colored black hole critical behavior
with opposite signs of the critical solution.
We find that as we tune $r_b$ closer and closer to the threshold,
the Reissner-Nordstr\"om solution corresponding to $w=0$ appears as a new 
\emph{approximate} unstable attractor before the colored black hole is 
approached (Fig.~\ref{f:w_evolns}); the term \emph{approximate} will be 
explained below.
Very close to $r_b^*$ this new attractor dominates
and the colored black hole attractor is no longer visible. 
(Seeing it would require an excessive amount of fine-tuning of $A_b$.)

We now analyze the behavior at the $r_b$-critical threshold in more detail.
First we look at the time the solution spends near the Reissner-Nordstr\"om
attractor in an $A_b$-critical search (the solid line in 
Fig.~\ref{f:Ab_timescaling}), this time with $r_b \approx r_b^*$.
Here we define $\tau$ as the time when $w$ first passes the value
$w=0.1$ at $\scri$ in slightly $A_b$-supercritical evolutions. 
Again this shows critical scaling, with a fitted exponent $\gamma = 10.67$.
This agrees well with the inverse of the dominant eigenvalue of the
Reissner-Nordstr\"om attractor (which has horizon radius $R_h = 2.57$):
$\lambda_0 = 0.09348 \, \Rightarrow \, 1/\lambda_0 = 10.70$.

\begin{figure}
\centerline{\includegraphics[width=.475\textwidth]{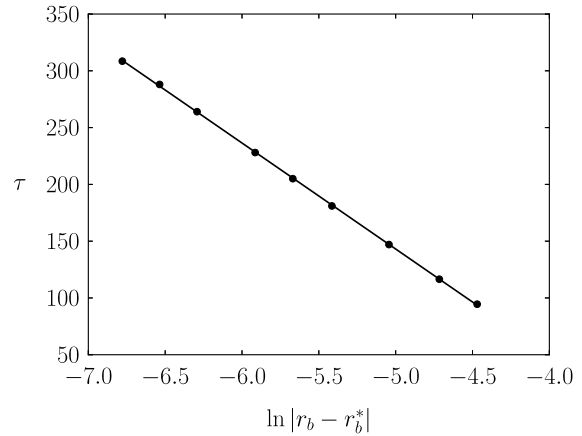}}
\caption{\label{f:rb_timescaling}
  The time $t=\tau$ when the evolution departs from the Reissner-Nordstr\"om 
  attractor as a function of $r_b$ along the critical line, 
  where $A_b$ has been tuned very closely to threshold.
  The line is a linear fit to the data points.
}
\end{figure}

Next we look at the departure time from the Reissner-Nordstr\"om attractor
(defined as above) \emph{tangential} to the critical line,
i.e. we tune $A_b$ very closely to threshold and vary $r_b$.
The result is shown in Fig.~\ref{f:rb_timescaling}.
Critical scaling with a fitted exponent $\gamma = 93.3$ is found,
which differs by about one order of magnitude from the exponent \emph{away}
from the critical line.
This value agrees well with the next-to-dominant eigenvalue of the 
Reissner-Nordstr\"om solution, 
$\lambda_1 = 0.01062 \, \Rightarrow \, 1/\lambda_1 = 94.2$.

\begin{figure}
\centerline{\includegraphics[width=.475\textwidth]{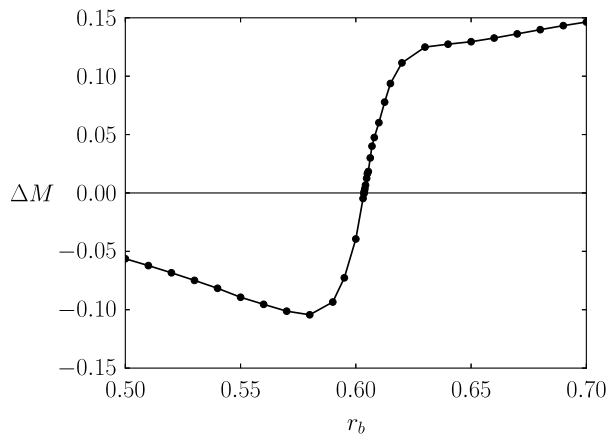}}
\caption{\label{f:rb_massgap}
  The mass gap $\Delta M = M_{f+} - M_{f-}$ across the $A_b$-critical
  threshold as a function of $r_b$.
}
\end{figure}

Finally we evaluate the mass gap $\Delta M = M_{f+} - M_{f-}$ between slightly
$A_b$-subcritical and $A_b$-supercritical evolutions as a function of $r_b$
(Fig.~\ref{f:rb_massgap}).
This vanishes linearly as $r_b \rightarrow r_b^*$, consistent with 
perturbations about Reissner-Nordstr\"om spacetime, which have vanishing
mass gap (Sec.~\ref{s:pertevoln}).
In fact, the zero of the mass gap allows for the most accurate determination 
of the critical value $r_b^*$.

All these results are consistent with the picture (Fig.~\ref{f:parameter_space})
that the codimension-two unstable attractor that divides the critical line in 
the two-dimensional parameter space into colored black hole solutions with 
opposite sign is the Reissner-Nordstr\"om solution.
Along its dominant unstable mode, the solution departs immediately to one of the
final Schwarzschild endstates.
Along its second unstable mode, it first approaches one of the copies of the
$n=1$ colored black hole (with either sign) before departing to Schwarzschild.

\begin{figure}
\centerline{\includegraphics[width=.475\textwidth]{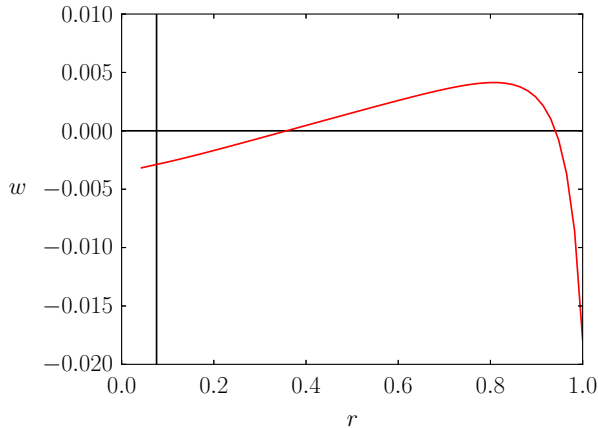}}
\caption{\label{f:w_closest}
  The Yang-Mills potential $w$ as a function of conformal radius $r$
  at time $t=250$ for an evolution with $r_b = 0.6036 \approx r_b^*$.
  Compare with $\delta w_2$ in Fig.~\ref{f:RN_modes}.
}
\end{figure}

There remains a puzzle though: the Reissner-Nordstr\"om solution has an 
infinite number of unstable modes.
By tuning two parameters in the initial data, one can eliminate the two 
dominant modes but in general not any of the other unstable modes.
These modes have an increasing number of radial oscillations.
When decomposing the near-critical solution into the eigenmodes, the
coefficients must therefore fall off fast with the mode number, as the
solution certainly remains smooth.
Thus we would expect the near-critical solution to look \emph{qualitatively}
like the third unstable mode, which has two zeros.
Indeed this is what we see (Fig.~\ref{f:w_closest}).
The third eigenvalue is much smaller than the second (by a factor $33$)
and hence the associated eigenmode is seen as almost constant in time.

The Reissner-Nordstr\"om solution is thus only an \emph{approximate}
codimension-two attractor, but one that is approached remarkably closely.

\section{Conclusions and discussion}
\label{s:concl}

We investigated static spherically symmetric black hole solutions to the
Einstein-Yang-Mills equations, in particular the $n=1$ colored black hole and
the magnetic Reissner-Nordstr\"om solution, and the role they play in
gravitational collapse.

A new formulation and numerical implementation of the equations on 
hyperboloidal surfaces of constant mean curvature (CMC) extending out to future 
null infinity was used \cite{Moncrief2009,Rinne2010,Rinne2013}.
This gives us access to the entire part of spacetime to the future of the 
initial hyperboloidal surface.
Our coordinates smoothly pass through the horizon to the interior of the 
black hole, allowing us to study black hole formation from regular initial data.

We began by constructing the relevant static solutions on CMC surfaces,
and we computed the unstable modes and corresponding eigenvalues 
in linear perturbation theory.
The modes are regular at the horizon.

We then studied nonlinear numerical evolutions of these linearly perturbed 
static black holes.
The mass gap between the final Schwarzschild black holes of evolutions with
either sign of the perturbation was computed for a range of horizon radii
for the first time and compared with the maximal mass gap that would result
if all of the black hole's hair either fell into the black hole or dispersed 
to infinity.
Even though obtaining accurate results for large horizon radii $R_h$ 
is numerically very challenging, our results suggest that the actual mass gap 
will likely only approach zero as $R_h \rightarrow \infty$.
Similarly, we computed the final mass of perturbed Reissner-Nordstr\"om black 
holes for a range of horizon radii.
In this case there is no mass gap.

For the Reissner-Nordstr\"om solution, we compared the dominant eigenvalue 
with a bound on the lifetime of unstable hairy black 
holes proposed by Hod \cite{Hod2008}.
This bound comes in two versions \eqref{e:hod}.
We find that the weak bound is satisfied for all values of the horizon radius.
If the quantity $\Delta \mathcal{E}$ appearing 
in the strong bound  is taken to be the \emph{entire} mass outside 
the horizon, as in \cite{Hod2008}, then this bound is violated 
for near-extremal Reissner-Nordstr\"om black holes.
If however $\Delta \mathcal{E}$ is taken to be the fraction of this mass that
actually falls into the black hole during the nonlinear evolution
(which is suggested by the derivation in \cite{Hod2008})
then the strong bound is satisfied and saturated very closely.
It is quite remarkable that the somewhat heuristic arguments from quantum 
information theory used in \cite{Hod2008} result in such a good prediction.

Finally we turned to critical behavior in gravitational collapse.
Our results confirm and demonstrate the universality of the phenomena first 
observed in \cite{Choptuik1999}
at the threshold between the two different final vacua ($w=\pm 1$) of the 
Yang-Mills field \emph{within} the class of evolutions that collapse to black 
holes (``Type III'' critical collapse).
The critical solution is the $n=1$ colored black hole, and the time spent
in its vicinity shows critical scaling with an exponent that is in agreement
with the unstable eigenvalue of the colored black hole.
We also showed that the mass gap between slightly subcritical and slightly 
supercritical evolutions agrees well with the mass gap observed by starting
off directly with the perturbed colored black hole critical solution.

The main result of this paper is a novel codimension-two critical phenomenon.
Using an extended family of regular initial data, we were able to probe
regions of parameter space where the colored black hole critical solutions
have \emph{opposite sign}. 
The existence of two copies of these solutions is a consequence of the 
invariance of the Einstein-Yang-Mills equations under $w \rightarrow -w$.
We constructed a two-parameter family of initial data that smoothly connects
these two regions in parameter space and investigated the boundary between them.
We gave strong evidence that the Reissner-Nordstr\"om solution appears as a new
codimension-two attractor.
In a neighborhood the evolutions show the expected critical
scaling of the time spent near the Reissner-Nordstr\"om solution,
consistent with the results from the linear mode analysis.
Along the dominant unstable mode, the solution departs immediately to one
of the two Schwarzschild endstates; along the subdominant mode, it first
moves towards one of the copies (with different signs of $w$)
of the colored black hole.
However, the Reissner-Nordstr\"om solution is only an approximate 
attractor because it has an infinite number of unstable modes, which cannot
all be tuned away using two parameters.
The contribution of the higher modes is remarkably small though, and at the 
time of closest approach the solution is dominated by the third eigenmode.
This appears to be the first time that a critical solution with an infinite
number of unstable modes was shown to play a role as an intermediate attractor
in gravitational collapse.

The fact that the Reissner-Nordstr\"om solution appears at the boundary
between $n=1$ colored black hole solutions of opposite sign came as a 
surprise---we expected this role to be taken by the $n=2$ colored black hole.
In two-parameter studies of Type II critical collapse in the 
five-dimensional vacuum Einstein equations \cite{Bizon2006}, the authors
found a new discretely self-similar solution with two unstable modes
as the codimension-two attractor, exploiting discrete symmetries similar
to our $w \rightarrow -w$ symmetry.
There are arguments that this behavior is quite generic \cite{Corlette2001}.
Of course we cannot exclude that there might be other families of initial data
that do have the $n=2$ colored black hole as a codimension-two attractor.

One might wonder whether similar behavior exists for
the standard Type I critical collapse separating black hole formation
and dispersal.
In this case the critical solution is the Bartnik-McKinnon 
soliton \cite{Choptuik1996}.
Indeed, by the same symmetry of the Einstein-Yang-Mills equations,
two copies of this solution with opposite signs of $w$ exist.
However, there is no smooth family of initial data covering regions
of critical collapse with both versions of the critical solution.
The reason is that regularity at the origin $r=0$ requires that the Yang-Mills
field be in one of its vacua $w=\pm 1$ at the origin at all times
(and our choice of $W$ as an evolved variable \eqref{e:wvsW} selects the 
$w=1$ vacuum).
It is impossible for the Yang-Mills field to switch from one vacuum to the other
at the origin during an evolution, and hence it is impossible to have both
solitons with opposite signs as critical attractors in the same smooth family 
of initial data.

Another interesting question suggested by our discovery of the codimension-two
critical behavior is whether superextremal ($M < 1$) magnetic 
Reissner-Nordstr\"om black holes can be formed in this process.
There is not much room for this because the minimum mass of the colored black 
holes along the critical line is $M\approx 0.83$, the mass of the 
Bartnik-McKinnon soliton.
We have not been able to construct a two-parameter family of initial data
that connects such sufficiently light colored black hole attractors of 
opposite sign without hitting regions of parameter space where the fields
\emph{disperse} instead of forming a black hole.
It could well be that this is the way cosmic censorship is enforced in this
context.

Finally we point out that we assumed a purely magnetic ansatz for the Yang-Mills
field \eqref{e:ymansatz}.
In \cite{Rinne2013} we considered the most general ansatz and used this in
numerical studies of power-law tails at future null infinity.
The question remains how the presence of a sphaleronic part of the Yang-Mills
field affects critical behavior;
this will be the subject of future work.


\acknowledgments 
I am particularly grateful to Piotr Bizo\'n for many enlightening discussions
and advice on this work.
Further thanks go to Vincent Moncrief, Shahar Hod, Lars Andersson,
Georgios Doulis and Christian Schell 
for helpful discussions and comments on the manuscript.
This research is supported by a Heisenberg Fellowship and grant RI 2246/2
of the German Research Foundation (DFG).


\bibliography{references}

\end{document}